\documentclass{article}

\PassOptionsToPackage{numbers, compress}{natbib}

\usepackage[utf8]{inputenc} 
\usepackage[T1]{fontenc}    
\usepackage{hyperref}       
\usepackage{url}            
\usepackage{booktabs}       
\usepackage{amsfonts}       
\usepackage{nicefrac}       
\usepackage{microtype}      
\usepackage{graphicx}
\usepackage{amsmath, amsthm}
\usepackage{amssymb}
\usepackage{caption}
\usepackage{subcaption}
\usepackage{color}
\usepackage{enumitem}
\usepackage[noend]{algpseudocode}

\usepackage[toc,page]{appendix}
\usepackage{float}
\usepackage{multirow}
\usepackage{longtable}

\usepackage{dirtytalk}
\usepackage{nopageno}

\usepackage[backend=biber,style=apa,natbib=true]{biblatex}
\DeclareLanguageMapping{english}{english-apa}

\def\inline#1:{\par\vskip 7pt\noindent{\bf #1:}\hskip 10pt}

\long\def\comment #1\commentend{}

\DeclareFontFamily{OT1}{pzc}{}
\DeclareFontShape{OT1}{pzc}{m}{it}{<-> s*pzcmi8t}{}
\DeclareMathAlphabet{\mathpzc}{OT1}{pzc}{m}{it}

\title{Publication Patterns' Changes due to the COVID-19 Pandemic: A longitudinal and short-term scientometric analysis}

\begin{document}



\author{Shir Aviv-Reuven \thanks{avivres@biu.ac.il} And Ariel Rosenfeld \thanks{ariel.rosenfeld@biu.ac.il}\\ Department of Information Sciences, Bar-Ilan University, Israel}




\maketitle
\begin{abstract}
In recent months the COVID-19 (also known as SARS-CoV-2 and Coronavirus) pandemic has spread throughout the world. In parallel, extensive scholarly research regarding various aspects of the pandemic has been published.
In this work, we analyse the changes in biomedical publishing patterns due to the pandemic. We study the changes in the volume of publications in both peer reviewed journals and preprint servers, average time to acceptance of papers submitted to biomedical journals, international (co-)authorship of these papers (expressed by diversity and volume), and the possible association between  journal metrics and said changes.
We study these possible changes using two approaches: a short-term analysis through which changes during the first six months of the outbreak are examined for both COVID-19 related papers and non-COVID-19 related papers; and a longitudinal approach through which changes are examined in comparison to the previous four years. 
Our results show that the pandemic has so far had a tremendous effect on all examined accounts of scholarly publications: A sharp increase in publication volume has been witnessed and it can be almost entirely attributed to the pandemic; a significantly faster mean time to acceptance for COVID-19 papers is apparent, and it has (partially) come at the expense of non-COVID-19 papers; and a significant reduction in international collaboration for COVID-19 papers has also been identified. 
As the pandemic continues to spread, these changes may cause a slow down in research in non-COVID-19 biomedical fields and bring about a lower rate of international collaboration.  

\end{abstract}


\section{Introduction}


The year 2020 began with the extremely fast spread of the COVID-19 pandemic and has already reached multiple peaks in the recent months \citep{WorldHealthOrganization, moore2020covid}.
Countries such as the US, India, Brazil and many others are struggling to flatten the curve.
The pandemic has impacted almost every aspect of life, ranging from the economy to tourism, political affairs, the arts and sports, thus there is a global effort in searching for ways to understand and cope with it.
These efforts lay, in great part, in the hands of the scientific research community. As such,  scholarly research and its publication patterns have also been greatly impacted by this current crisis. 

The volume of COVID-19 related publications, especially in the biomedical fields, has increased and has continued to sharply increase since January 2020.  However, apart from this increase in volume, other changes in scholarly research are also taking place.
Many journals and publication databases now allow free access to COVID-19 related articles and data (\textcolor{blue}{\href{https://coronavirus.elsevier.com/}{Elesevier coronavirus Research hub}, \href{https://www.thelancet.com/coronavirus}{The Lancet}}). Data sets  of these articles such as the \textcolor{blue}{\href{https://www.semanticscholar.org/cord19}{CORD-19}} have also been curated for the creation of analysis tools to aid in the fight against this disease.

While it is clear that scholarly publication patterns have changed dramatically due to the pandemic, 
it remains unclear how these changes are manifested in a few key aspects. We focus on four such aspects by setting the following research questions:
\begin{enumerate}
    \item \textit{How has the volume of scholarly literature in preprint servers and journals changed due to the pandemic? Specifically, we hypothesize that the focus on COVID-19 caused a reduction in volume of publications of other, non-COVID-19 papers in the same venues.}
    \item \textit{How are COVID-19 publications in journals related to their associated metrics? Specifically, do journals with higher scientometric scores publish more COVID-19 related papers then journals with lower scores?}
    \item \textit{How quickly are COVID-19 and non-COVID-19 papers accepted for publication? The peer review process of journals is usually slow, but currently there is a need for a fast turn around, especially for COVID-19 papers. Specifically, we hypothesise that in order to cope with this need for fast turn around, the time until the acceptance of COVID-19 papers has been reduced significantly from \say{normal} acceptance time and that the time until the acceptance of non-COVID-19 papers has slowed down in order to facilitate that.}
    \item \textit{How has international publication and (co-)authorship changed? We hypothesise that the pandemic has caused a significant increase in international collaboration. We set to analyse international collaboration from two unique axes: 1) The diversity of countries which collaborate with one another; and 2) The number of internationally co-authored COVID-19 publications.} 
\end{enumerate}
To address these questions we employ both a short-term analysis technique, focusing on the first 6 months of 2020 (the first six months of the outbreak), and a longitudinal analysis technique through which we compare the publication patterns across the last five years (2016-2020). Our study employs a set of statistical tests in order to ascertain statistically significant changes. These tests are conducted at both the short-term and longitudinal levels. At the short-term level, these tests indicate if any statistically significant differences exist when comparing COVID-19 papers to non-COVID-19 ones. At the longitudinal level, these tests indicate if any statistically significant differences exist when comparing papers published prior to the pandemic to those published during the pandemic.
We focus on two main types of venues for research publication: peer reviewed Journals and Preprint servers. Preprint servers are becoming widely used in other fields of research, such as Computer Sciences and Physics, but up until the pandemic the usage of such publication venues in the biomedical fields was limited \citep{desjardins2013case, maslove2018medical}.    


Understanding changes in publication patterns during the pandemic is valuable due to the possible implications. As the pandemic does not seem to be coming to a stop, these changes, for good and for bad, may have prolonged effects that should be considered by journal editors, recruiting and promotion committees, funding agencies and others. 


This paper is organized as follows: Section \ref{sec:Related_work} presents the background and related work in scientometric analysis of pandemic related research. In Section \ref{sec:Methodology} we describe the data and tools used in this study. Section \ref{sec:Results} presents the results to the research questions we posed. We conclude the paper with a discussion in Section \ref{sec:Conclusion}.





\section{Background and Related Work}
\label{sec:Related_work}

Numerous scientometric studies have examined how publication patterns vary during or following a pandemic (see  \citet{zhang2020scientific} and references therein). These studies commonly focus on one or a few aspects of scientometrics such as growth of publications in various databases, research funding agencies' countries, average time to acceptance and international collaboration patterns. 
In these works, two standard techniques are often used: a short-term technique in which publication pattern changes are analysed during a pandemic and a longitudinal technique in which the publication pattern analysis is focused on a collection of papers related to viral pandemics, written over a long period, usually several years.  

Recent studies on the COVID-19 pandemic follow these two techniques as well. Adopting the short-term analysis technique,  \citet{da2020publishing} have examined publication volumes of COVID-19 papers and identified top journals, countries and authors. Similarly, \citet{costa2020scientific} performed keyword analysis and identified the most productive countries, institutions, authors and journals, \citet{lou2020coronavirus} observed publication types, journals and publication countries, and \citet{gianola2020characteristics} identified that during the first five months of the pandemic most of the COVID-19 scientific literature comprised of short reports, opinions and perspectives. \citet{chahrour2020bibliometric} focused on the international distribution of COVID-19 publications compared with the number of COVID-19 cases in the respective countries, again adopting a short-term analysis approach. Common to the above studies is the focus on COVID-19 related publications. These studies do not consider the possible changes in publication patterns of papers unrelated to the COVID-19 pandemic which were published during the pandemic. One exception is \citet{homolak2020preliminary} who do focus on both COVID-19 related papers and non-COVID-19 related papers in their short-term analysis observing time to publication, authorship and affiliation counts.

Adopting a longitudinal approach, \citet{kun2020time} observed the extremely short time to acceptance for COVID-19 papers. The author focused on the first three months of 2020 for COVID-19 papers and compared them to papers on other corona viruses. 
\citet{ahmad2020identifying} have taken a different approach, focusing solely on COVID-19 papers yet examining them over the years 2011-2020. Similarly, \citet{tao2020covid, mao2020status,zhai2020research, malik2020scientometric} studied the same for the years 2000-2020. These studies examined the publications' countries of origin, collaboration networks, authors, keywords and additional publication characteristics. 
\citet{malik2020scientometric} found that the number of nCOV related research papers has spiked several times in the last two decades, correlating with the post SARS and MERS pandemics. In the same vein, \citet{kagan2020scientometric} analysed publications related to multiple nCov viruses and compared those to influenza and additional viruses, and \citet{zhang2020scientific} did a comparative bibliometric study of multiple outbreaks and performed a preliminary analysis of the COVID-19 outbreak. 
Other studies have performed both longitudinal and short-term analyses of nCov papers in which they examined international collaboration \citep{lee2020Scientific, cai2020international}. Their studies show that countries affected more by the virus as well as those with higher GDP tended to publish more in international collaborations, and that team sizes for COVID-19 papers dropped during the first months of pandemic as well the number of papers published in international collaborations.

Our study further compares preprint servers and peer-reviewed journals as possible dissemination venues for COVID-19 research output. 
Prior research by \citet{krumholz2020preprints} and \citet{johansson2018preprints} have shown an increase in the usage of preprint servers in previous pandemics. Recently, evidence was provided to support their findings in the current pandemic, as well \citep{fraser2020preprinting,fry2020consolidation}. \citet{torres2020daily} have also identified this growth observing eight different repositories. They further observed the total growth in volume, showing that the number of COVID-19 papers produced doubles every 15 days.
The work by \citet{vasconcelos2020modelling} showed an overall growth of preprint papers in  repositories across multiple fields and modeled this growth.
Our study complements the above in several respects: 1) By providing both short-term and longitudinal statistical analysis, our study gives a wider perspective in which the influence of the pandemic on current research can be observed; 2) We focus on both COVID-19 and non-COVID-19 papers, providing an assessment of how the effects of the pandemic differ in respect to these two types of papers; 3) Our study analyzes time to acceptance for COVID-19 and non COVID-19 publications as well as international collaboration by the diversity of the countries that are collaborating. These aspects have been minimally researched in pandemics in general and during the COVID-19 pandemic in particular. 4) Our study examines two types of venues, namely Preprint servers and scholarly journals; and 5) To the best of our knowledge, we provide the most extensive scientometric-based research on COVID-19 publications to date.

\section{Methodology}  
\label{sec:Methodology}
\subsection{Sources}
\label{subsec:Sources}
The data for this research was obtained from four main sources:
\begin{itemize}
    \item Elseviers' ScienceDirect\footnote{\textcolor{blue}{\href{https://www.sciencedirect.com}{https://www.sciencedirect.com}}} and Scopus\footnote{\textcolor{blue}{\href{https://www.scopus.com}{https://www.scopus.com}}}. ScienceDirect is a full-text scientific database which is part of SciVerse. Scopus is an abstract and citation database of peer-reviewed literature. Utilizing both ScienceDirect and Scopus API we extracted data on COVID-19 related journal articles. 
    \item medRxiv\footnote{\textcolor{blue}{\href{https://www.medrxiv.org}{https://www.medrxiv.org}}} (pronounced \say{med-archive}) is a free online archive and distribution server for complete but unpublished manuscripts (preprints) in the medical, clinical, and related health sciences. The server was founded by Cold Spring Harbor Laboratory (CSHL), a not-for-profit research and educational institution, Yale University, and BMJ (mostly referred to as the British Medical Journal), a global healthcare knowledge provider.
    \item bioRxiv\footnote{\textcolor{blue}{\href{https://www.biorxiv.org}{https://www.biorxiv.org}}} (pronounced \say{bio-archive}) is a free online archive and distribution service for unpublished preprints in the life sciences. It is operated by CSHL. 
    \item arXiv\footnote{\textcolor{blue}{\href{https://arxiv.org}{https://arxiv.org}}} is an open archive for scholarly preprints in various fields. It is maintained and operated by Cornell University. 
\end{itemize}

In addition to the above described datasets, we have also extracted supplementary data from PubMed which is a web-portal of the medical database MEDLINE; and ScimagoJR - a publicly available portal which includes journals' and countries' scientific indicators developed from the information contained in the Scopus database, created by SCImago research group \citep{gonzalez2010new}.
The selection of ScienceDirect and Scopus was due to their wide indexing of journals as well as their API for data extraction. medRxiv and bioRxiv  were selected due to their specialization in the biomedical research fields. Similarly, the arXiv repository was chosen due to its high usage across multiple fields.  As the arXiv server is used for many fields of research and our study focused on biomedical papers, the data extraction from the arXiv server was limited only to the \say{quantitative biology} field.
\subsection{Retrieval Process}
\label{subsec:RetrievalProcess}
To understand the publication behaviour in the first months of the pandemic, we analysed the data which was extracted from the sources described in Section \ref{subsec:Sources}.
The search in ScienceDirect was done using the search query \textit{\say{COVID-19} OR Coronavirus OR \say{Corona virus} OR Coronaviruses OR \say{2019-nCoV}} everywhere in the document.
The biomedical journals with the highest numbers of COVID-19 related publications were selected for further analysis. For each journal, data for all COVID-19 and non-COVID-19 related papers was downloaded separately. In the same manner, data for all papers in each year of our analysis (2016-2020) was downloaded and split up by months according to the online availability date of the paper. Additional data for each paper was extracted via its DOI from PubMed Entrez and through the Scopus API. This included the dates the paper was received by the journal, accepted for publication and available online, the authors' affiliations and countries, the journals' urls and the associated scientometrics. We focus on the \say{Scimago Journal Rank} (SJR) \citep{gonzalez2010new} metric as our data was collected from Scopus. Duplicate papers, identified by their DOIs, as well as papers with missing data (DOI, authors or countries) were removed automatically. Additional papers which were removed were those with inaccurate dates or insufficient date information as described in Section \ref{subsec:AnalysisApproach}.  The total number of papers excluded from our analysis was 347 out of 7419. The average yearly percentage of papers removed from analysis was $3.76\%$ (for the years 2016-2020, inclusive). Records for the examined papers were analysed according to various attributes including publicizing journal, authors, publicizing countries and dates.

In order to retrieve and analyse COVID-19 data from bioRxiv, medRxiv and arXiv, we queried the archive servers with the search query: \textit{\say{COVID-19} OR Coronavirus OR \say{Corona virus} OR Coronaviruses OR \say{2019-nCoV}}. This query was executed separately for each of the first six months in 2020. To perform the longitudinal aspect of our analysis we further queried the repositories for all papers published in each of the first six months of 2016 to 2020 separately.
The retrieved results were downloaded and automatically analysed using designated scripts written by the authors.

A subset of the results was manually tested to ensure both accuracy of the data and the scripts. A random subset of several dozen articles was chosen and manually examined by the authors. The relevant dates, authors, countries and additional data were compared against the data automatically extracted by the scripts to ensure no mismatch. No discrepancies were found in this manual verification.

\subsection{Analysis Approach}
\label{subsec:AnalysisApproach}
We analyse four main aspects of our data as pertaining to our posed research questions: 1)  Publication growth that has occurred during the pandemic and, specifically, the venues which have contributed to this growth; 2) Impact of journals' Scientometric indicators on publication behaviour; 3) Changes in the time to acceptance of peer reviewed papers; and 4) Changes in authors' countries of affiliation and international collaborations. 
We define the time to acceptance as the period between the \say{date received} and the \say{date accepted} or the \say{date online} for a paper, whichever is earlier. Data entries for which the \say{date received} or both the \say{date accepted} and the \say{date online} were missing or corrupted were omitted from our analysis as well as inaccurate data entries for which the \say{date received} was later than (or the same as) the \say{date online} or the \say{date accepted}.
The cleaning methodology is detailed in Section \ref{subsec:RetrievalProcess}
To conduct our collaboration analysis we define international collaboration as papers authored by two or more authors affiliated with institutions in different countries. 
We examine two facets of international collaborations: 
\begin{itemize}
    \item \textit{Diversity of collaboration}, i.e., the number of countries with which each country has collaborated over a given time period.
For example, a country which has published papers with 10 other countries is more internationally collaboratively diversified than a country who has published with 5 other countries, irrespective of the number of papers published.
    \item \textit{\say{volume} of collaboration}, i.e., the number of publications which each country has published in collaboration with other countries over a given time period. For example, a country who has published 10 papers with one other country is more internationally productive than a country who has published 5 papers, even if each of the papers is published with a different country.
\end{itemize}

The four aspects are analysed and reported for both COVID-19 papers as well as \say{standard} non-COVID-19 papers published during the pandemic. This comparative approach allows us to identify the effects of the pandemic on the examined aspects both for pandemic-related research as well as the standard biomedical research published during that time.   
The above aspects are further analysed for pre-pandemic papers published in the years 2016-2019 and compared against the COVID-19 papers and non-COVID-19 papers published in 2020. Our analysis was performed separately for COVID-19 papers and for non-COVID-19 papers in each of the first six months of 2020 and repeated for each of the first six months of each of the previous four years in our research. This part of the study allows us to identify the possible effects of the pandemic on the examined aspects in a longitudinal view. Some of the following analyses, especially regarding time to acceptance, authors' country of affiliation and international collaboration, require a large volume of publications. Thus, for these analyses we selected a subset of biomedical journals with the highest number of COVID-19 paper publications. These are shown in Table \ref{tbl:cov_journals_growth}. In order to select these journals we performed the queries described in Section \ref{subsec:RetrievalProcess} and ordered them by their number of COVID-19 related publications. From the journals with the highest numbers of COVID-19 publications we selected journals in biomedical fields according to their associated categories in Scopus.

All data and code is available under
\textcolor{blue}{\href{http://www.github.com/shirAviv/covid-19-scientific-papers} {www.github.com/shirAviv/covid-19-scientific-papers.}}

\section{Results}
\label{sec:Results}

\subsection{Publication Growth}\label{sec:pubG}

Analysing the publication growth in the first six months of 2020 shows that not only has there been a huge surge of COVID-19 related publications, as one could expect, but also that these publications are disseminated across multiple venue types. As discussed before, this work focuses on two venue types- preprint servers and peer reviewed journals.

\subsubsection{Publication Growth in Preprint Severs}
\label{subsub:pub_growth_servers}

\begin{figure}
	    \includegraphics[width=\textwidth]{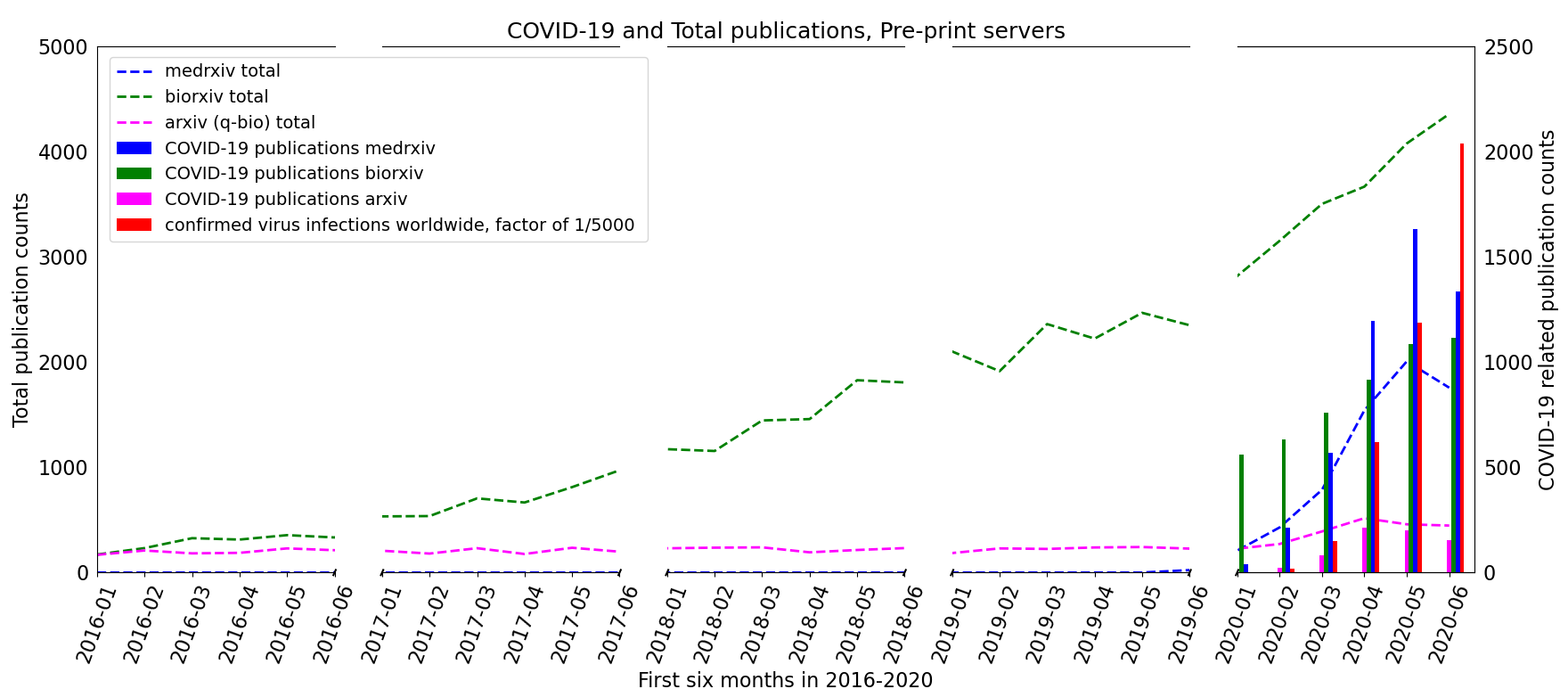}
        \caption{Publication growth in the examined preprint repositories during the first six months of the years 2016-2020. Dotted plots - Total papers published. Bars - COVID-19 related papers, compared with the pandemic spread (factor of 1/5000).}
        \label{fig:Publication_growth_cov_totals_archives}
\end{figure}

Figure \ref{fig:Publication_growth_cov_totals_archives} shows that preprint servers are considered a legitimate and even valuable source of dissemination at this time. Interestingly, the publication growth was observed for both COVID-19 publications as well as for non-COVID-19 related papers in all of the preprint servers we analysed. Furthermore, the growth in COVID-19 publications increased in a similar fashion to the international spread of the pandemic, although some decline was apparent in June in the medRxiv preprint server.

The sharpest increase in publications was observed in medRxiv, which was created as a preprint server in mid-2019. While the total number of papers published in it has increased from 200 in January 2020 to nearly 1800 in June 2020 (a factor of 9), the number of COVID-19 related papers increased from 40 in January 2020 to 1350 in June 2020 (a factor of 34). COVID-19 papers showed an increase in percentage from 19.5\% to 75\% of total papers published on the server during these months.

Focusing on the arXiv preprint server, which is well known for publications in Physics and Computer Science, reveals that the number of publications in biology (quantitative-biology field in the arXiv, q-bio for short) was low over the previous four years with a slow, almost flat, increase. However, from January 2020 until June 2020, the total number of publications in the q-bio field had almost doubled and the number of COVID-19 related papers had increased by a factor of 8. The percentage of COVID-19 papers increased from 20\% in the month of February to 35\% in June, of total papers published in the q-bio field of the server in these months\footnote{No COVID-19 related papers were published in the arXiv q-bio field in January.}.

\subsubsection{Publication Growth in Journals}
\label{subsub:pub_growth_journals}
Turning to the analysis of peer reviewed journal publications, 
we observe the growth of COVID-19 related papers in this venue. Table \ref{tbl:cov_journals_growth} focuses on the top COVID-19 publishing journals in which, especially in April, May and June 2020, COVID-19 papers comprised a substantial percentage of papers published in these journals. Journals in the table are ordered according to their SJR score from 2019. 

\begin{figure}
	    \includegraphics[width=\textwidth]{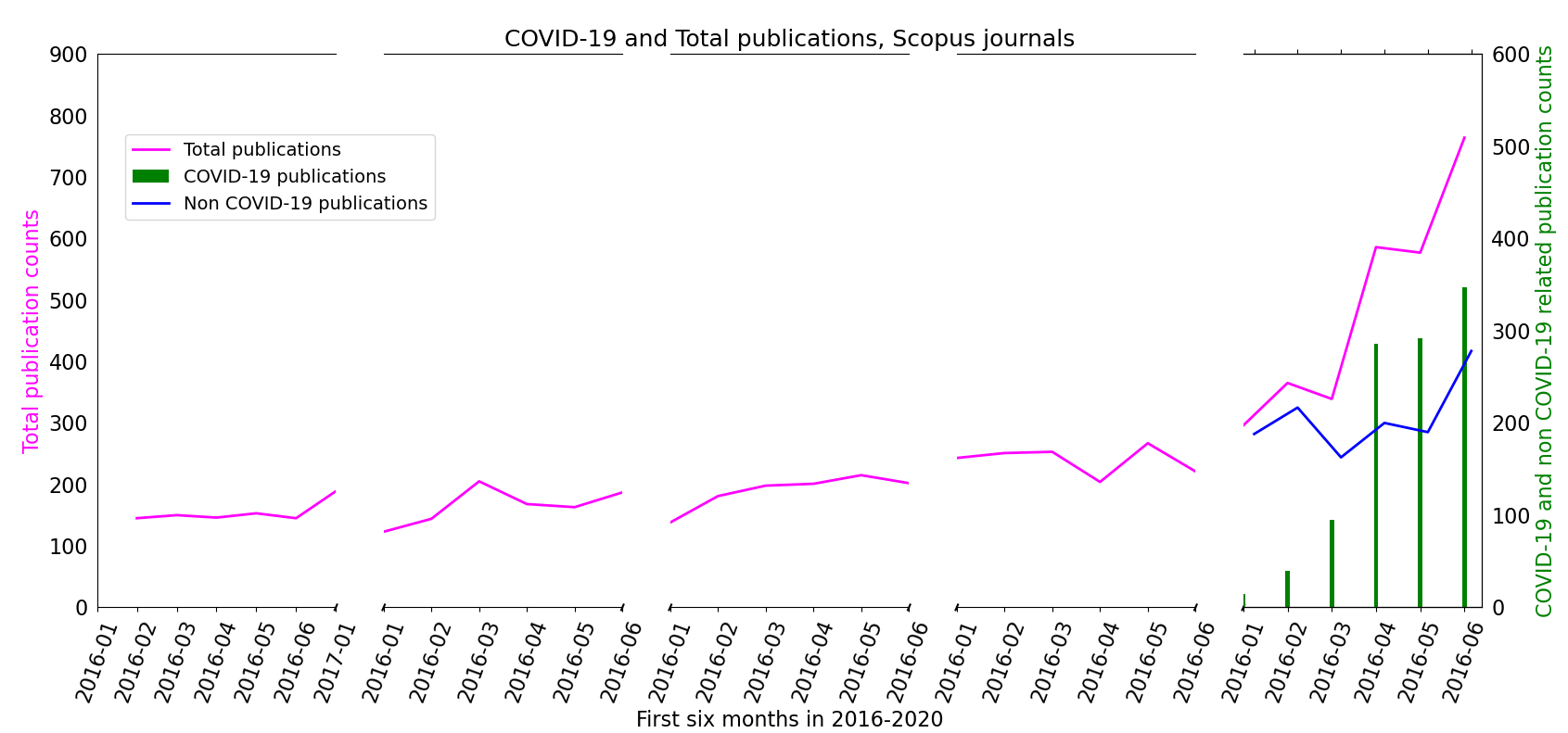}
        \caption{Publication growth in the examined Scopus journals during the first six months of the years 2016-2020. Plots - Total papers published (Magenta) and non COVID-19 related papers (Blue). Bars (Green) - COVID-19 related papers. (The list of journals is shown in Table  \ref{tbl:cov_journals_growth}).}
        \label{fig:Publication_growth_cov_totals}
\end{figure}

We further analysed the growth from a longitudinal aspect, observing the first six months of the years 2016-2020.
Similar to Figure \ref{fig:Publication_growth_cov_totals_archives} which presents the growth in preprint papers, Figure \ref{fig:Publication_growth_cov_totals} shows the COVID-19 publication growth for journals in Table \ref{tbl:cov_journals_growth} as compared with publication growth over the last five years. While we can see a large surge in the total number of publications as compared to previous years, COVID-19 publications seem to account for virtually the  entire growth. Specifically, the number of non-COVID-19 related publications follows the same pattern of previous years.

\setlength{\tabcolsep}{0.3pt}

\begin{center}
\small
\begin{longtable}[ht]{c|c|cc|cc|cc|cc|cc|cc|}
\caption[COVID-19 papers growth]{COVID-19 papers' growth in Scopus journals in the first six months of 2020. For each journal in each month: Cov19 - COVID-19 related papers; Total - total published papers; \% - percent of COVID-19 papers. The table shows the selected subset of journals ordered by their SJR score.}
\label{tbl:cov_journals_growth} \\

%

    \multirow{2}{*}{Journal} &
    \multirow{2}{*}{SJR} &
    \multicolumn{2}{c}{January} &
    \multicolumn{2}{c}{February} &
    \multicolumn{2}{c}{March} &
    \multicolumn{2}{c}{April} &
    \multicolumn{2}{c}{May} &
    \multicolumn{2}{c}{June}\\
& & Cov19 & Total &  Cov19 & Total &  Cov19 & Total &  Cov19 & Total &  Cov19 & Total &  Cov19 & Total  \\ \hline
    \hline
\endfirsthead
\multicolumn{3}{@{}l}{\ldots continued}\\\hline
\multirow{2}{*}{Journal} &
    \multirow{2}{*}{SJR} &
    \multicolumn{2}{c}{January} &
    \multicolumn{2}{c}{February} &
    \multicolumn{2}{c}{March} &
    \multicolumn{2}{c}{April} &
    \multicolumn{2}{c}{May} &
    \multicolumn{2}{c}{June}\\
& & Cov19 & Total &  Cov19 & Total &  Cov19 & Total &  Cov19 & Total &  Cov19 & Total &  Cov19 & Total  \\ \hline
    \hline
    \endhead
\multirow{2}{*}{The Lancet}\\{Infectious Diseases}                     &  9.040 &              0 &     38.0 &              6 &     48.0 &             22 &     66.0 &             32 &     71.0 &             25 &     69.0 &             28 &    120.0  \\
&&& 0.00\% & & 12.50\% & &     33.33\% & &     45.07\% & &     36.23\% & &     23.33\% \\
\hline
\multirow{2}{*}{The Lancet}\\{Global Health}                           &  8.055 &              0 &     34.0 &              1 &     38.0 &             10 &     37.0 &             18 &     40.0 &             11 &     31.0 &              6 &     32.0   \\
&&& 0.00\% & &   2.63\% & &    27.03\% & &    45.00\% & &    35.48\% & &     18.75\% \\
\hline
\multirow{2}{*}{The Lancet}\\{Respiratory Medicine}                    &  7.516 &              0 &     22.0 &              2 &     36.0 &              9 &     38.0 &             13 &     27.0 &             17 &     30.0 &             16 &     49.0  \\
&&& 0.00\% & &      5.56\% & &     23.68\% & &    48.15\% & &     56.67\% & &     32.65\% \\
\hline
\multirow{2}{*}{International Journal}\\{of Infectious Diseases}       &  1.437 &              2 &     53.0 &              6 &     63.0 &              7 &     42.0 &             28 &     79.0 &             31 &     82.0 &             28 &     99.0 \\ 
&&& 3.77\% &&    9.52\% &&   16.67\% &&    35.44\% &&     37.80\% &&     28.28\% \\
\hline
\multirow{2}{*}{Journal of}\\ {Hospital Infection}              &  1.295 &              2 &     15.0 &              5 &     22.0 &             10 &     23.0 &             11 &     32.0 
&        16 &     33.0 &             17 &     37.0 \\
&&&     13.33\% &&    22.73\% &&   43.48\% &&     34.38\% &&     48.48\% &&     45.95\% \\
\hline
\multirow{2}{*}{Travel Medicine}\\ {and Infectious Disease}        &  1.075 &              5 &     18.0 &              9 &     13.0 &             17 &     26.0 &             39 &     56.0 &             19 &     31.0 &             21 &     31.0 \\
&&&     27.78\% &&     69.23\% &&     65.38\% &&     69.64\% &&     61.29\% &&    67.74\% \\
\hline
\multirow{2}{*}{American Journal}\\{of Infection Control}              &  0.989 &              1 &     22.0 &              0 &     17.0 &              0 &     19.0 &             10 &     31.0 &             13 &     28.0 &             22 &     46.0 \\
&&&      4.55\% &&      0.00\% &&      0.00\% &&     32.26\% &&     46.43\% &&    47.83\% \\
\hline
\multirow{2}{*}{}\\{Journal of Infection}                               &  0.989 &              3 &     24.0 &              9 &     25.0 &             12 &     19.0 &             51 &     59.0 &             46 &     58.0 &             75 &     96.0 \\
&&&     12.50\% &&     36.00\% &&     63.16\% &&     86.44\% &&     79.31\% &&    78.12\% \\
\hline
\multirow{2}{*}{Asian Journal}\\{ of Psychiatry}                        &  0.736 &              0 &      5.0 &              1 &     53.0 &              3 &     16.0 &             38 &     82.0 &             18 &     53.0 &             46 &     96.0\\
&&&      0.00\% &&      1.89\% &&     18.75\% &&     46.34\% &&     33.96\% &&    47.92\% \\
\hline
\multirow{2}{*}{Diabetes \& Metabolic}\\{Syndrome: Clinical...} &  0.124 &              0 &     19.0 &              0 &     20.0 &              2 &     10.0 &             13 &     36.0 &             29 &     59.0 &             36 &     60.0\\
&&&      0.00\% &&     0.00\% &&     20.00\% &&     36.11\% &&     49.15\% &&     60.00\% \\
\hline
\multirow{2}{*}{Medical Hypotheses}           &  0.108 &              1 &     46.0 &              1 &     30.0 &              3 &     43.0 &             33 &     73.0 &             67 &    103.0 &             52 &     98.0 \\
&&&     2.17\% &&     3.33\% &&      6.98\% &&     45.21\% &&     65.05\% &&     53.06\% \\
\hline
\hline
\end{longtable}
\end{center}

Turning to the journals' SJR metric, the results show that the growth in COVID-19 related publications is correlated with the journals' SJR. In low ranking journals, hardly any COVID-19 papers were published. In contrast, the vast majority of COVID-19 papers were published in the highest ranked journals. Specifically, using the SJR score as a sorting criteria, 31\% of the COVID-19 papers we analysed were published in top ranking journals ($\sim$20\% of the analysed journals) with SJR ranking between 7.516 and 14.55. Only 9.8\% of the COVID-19 papers were published in bottom ranking journals ($\sim$20\% of analysed journals) with SJR ranking between 0.103 and 0.11. The number of articles used in our analysis was aggregated in each journal over the first six months of the pandemic. Table \ref{tbl:cov_sjr} shows the top ranking journals and the lowest ranking journals with the number of COVID-19 articles they published. Excluding two low ranked journals with an exceptionally high number of COVID-19 publications, only 1.8\% of the COVID-19 papers were published in bottom ranking journals ($\sim$20\% of the analysed journals) with SJR ranking lower than 0.25.

\begin{table}
\centering
\caption{COVID-19 papers published in journals with high and low SJR ranking. \% - percent of COVID-19 papers per journal.}
\label{tbl:cov_sjr} 

\begin{tabular}{lrrr}

\toprule
Journal name &  COVID-19 papers &     SJR &    \%  \\

\midrule
The Lancet                                         &            344.0 &  14.550 &  19.38 \\
The Lancet Infectious Diseases                     &            113.0 &   9.040 &   6.37 \\
The Lancet Global Health                           &             46.0 &   8.055 &   2.59 \\
The Lancet Respiratory Medicine                    &             57.0 &   7.516 &   3.21 \\
\hline
\hline
Brazilian Journal of Anesthesiology                &              7.0 &   0.110 &   0.39 \\
Visual Journal of Emergency Medicine               &              7.0 &   0.110 &   0.39 \\
Medical Hypotheses                                 &            157.0 &   0.108 &   8.84 \\
New Scientist                                      &              3.0 &   0.103 &   0.16 \\
\bottomrule
\end{tabular}
\end{table}

The Pearson correlation between the total number of COVID-19 publications and the SJR of the associated journal is moderately positive with $r=0.57$, and statistically significant at $p=0.0053$.
We further calculated the Pearson correlation between the percent of COVID-19 papers out of total papers, summed over the first six months of 2020, for each journal and the SJR score of that journal. The correlation is positive with $r=0.177$ but not statistically significant at $p=0.43$. This result could be due to the large amount of papers published in most high ranking journals irrespective of COVID-19.



\subsection{Time to Acceptance}
\label{subsec:timeToAcceptance}

Following the results reported in Section \ref{subsub:pub_growth_journals}, 
we now turn to investigate how this publication growth has affected the time to acceptance for the papers. Naturally, time to acceptance is only applicable to journal publications and not for preprint ones.
For each of the journals analysed in Table \ref{tbl:cov_journals_growth}, we calculated the time to acceptance in each of the six examined months for the years 2016-2020 by calculating mean and standard deviation (std) of the time to acceptance. 

\begin{figure}
	    \includegraphics[width=\textwidth]{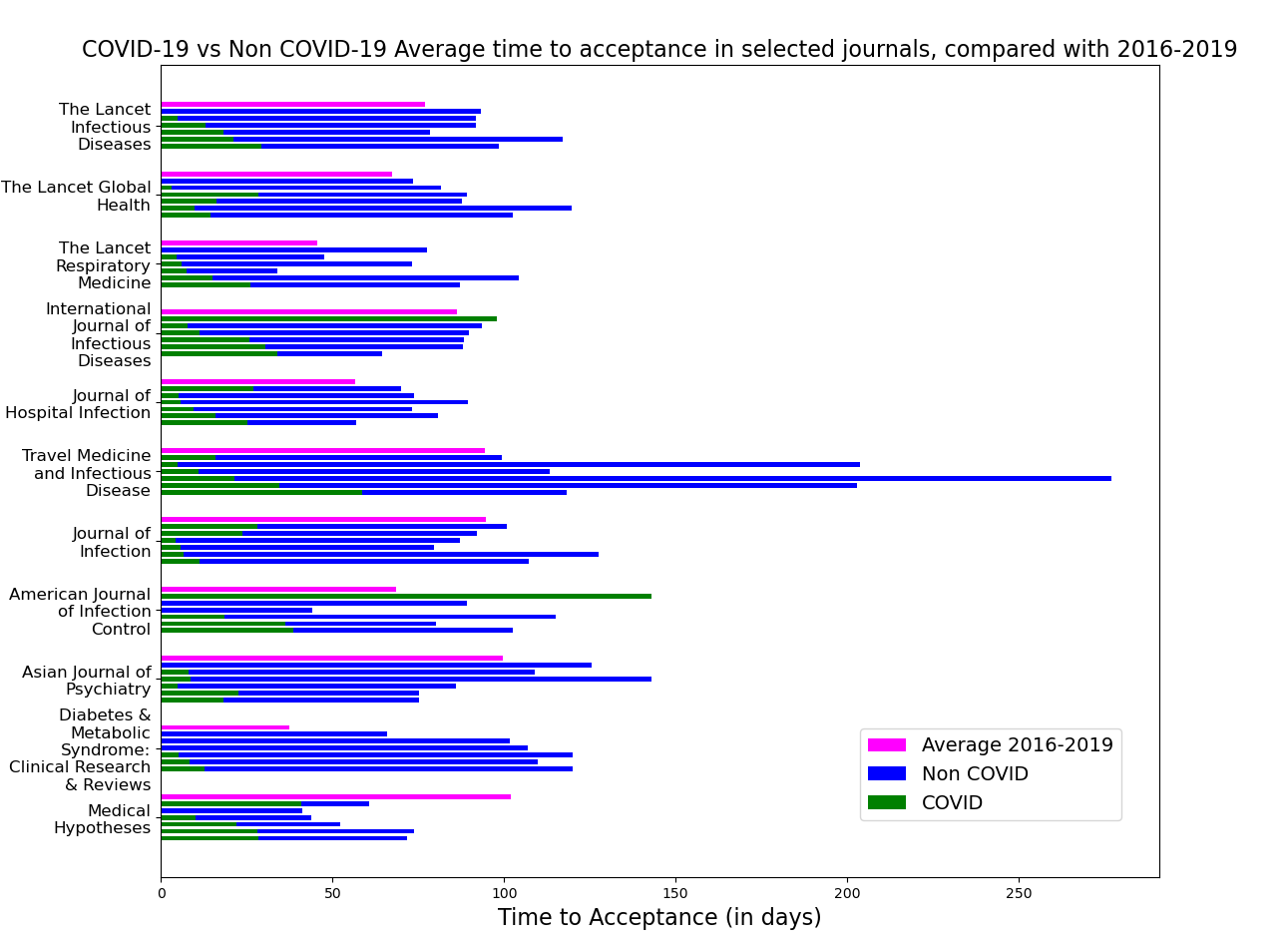}
        \caption{Mean time to acceptance in examined Scopus journals during the first six months of the year 2020 and the average of the first six months of 2016-2019. For each journal - Average in preceding years (magenta), COVID-19 (green), non-COVID-19 (blue). From top: preceding years' average, January, February, March, April, May and June (Bottom bar).} Journals are ordered according to SJR score, as shown in Table \ref{tbl:cov_journals_growth}.
        \label{fig:cov-19-non_cov-acceptance}

\end{figure}

Figure \ref{fig:cov-19-non_cov-acceptance} displays the mean time to acceptance for COVID-19 and non-COVID-19 publications in each of the first six months in 2020 alongside the mean time to acceptance of all papers published in the same months in 2016-2019, inclusive.
The journals in the figure are ordered by their rank. Observing the mean time to acceptance for COVID-19 papers in all journals, starting from the month of February 2020 onward, we see that it is extremely short, both when compared to non-COVID-19 papers from the same month and onward and when compared to papers from the previous years. 

We performed a series of one tailed t-tests comparing mean time to acceptance of COVID-19 papers to non-COVID-19 papers. The test was performed for each of the months February-June of 2020\footnote{Data for January was insufficient to perform this test.} as well as for the average time to acceptance for aggregated papers in the months January-June of 2020. Values are shown in Table \ref{tbl:acc_time_stats_cov_non_cov}. As can be seen, all of these tests revealed a statistically significant difference in mean time to acceptance with $p<0.05$. In February 2020, for example, the average time to acceptance of non-COVID-19 papers was almost 10 times longer than that of COVID-19 papers. Across the examined period, on average, non-COVID-19 papers experienced an average time to acceptance of 91.3 days compared to 19.3 days for COVID-19 papers (a factor of 4.7). 

Additional support for the above phenomena can be seen in Table \ref{tbl:acc_time_stats_history}. Taking a longitudinal approach, we compare each pair of consecutive years as to the mean time to the acceptance of papers from our examined journals. As can be seen in the table, up to and including 2019, no significant changes were observed. For 2019 and 2020 we see that papers published in 2020 experienced a longer time to acceptance, both when considering the entire publication set as well as when focusing on non-COVID-19 papers alone. 

Due to our previous finding that high ranked journals yield a higher number of COVID-19 papers, we expected this to affect the time to acceptance as well. Two contradicting hypotheses can be assumed: 1) The high volume of papers \textit{submitted} to high ranked journals yields a longer acceptance time; and
2) The high volume of \textit{accepted} papers to high ranked journals implies a faster review process and a shorter acceptance time.

In order to examine these hypotheses, we calculated the Pearson correlation between the SJR score of a journal and the mean acceptance time of COVID-19 related papers for that journal. The correlation was found to be weakly negative, as would be expected from the second hypothesis, but not statistically significant ($r=-0.266$, $p=0.43$). 

\setlength{\tabcolsep}{0.9pt}
\small
\begin{table}
\begin{center}
 \caption{Time to acceptance for COVID-19 and non-COVID-19 papers in the first months of the pandemic. Mean, std, \textit{T-Test  statistic} and p-value. Results in bold are statistically significant.}
    \label{tbl:acc_time_stats_cov_non_cov}
    \begin{tabular}{c|cc|cc|c|c}
    \toprule
    \multirow{2}{*}{Months} &
    \multicolumn{2}{c}{COVID-19} &
    \multicolumn{2}{c}{non-COVID-19} &
    \multirow{2}{*}{t-test statistic} &
    \multirow{2}{*}{p-value} \\
& Mean & std & Mean & std & & \\
   \midrule
February & 9.71 & 26.72 & 88.15 & 80.3 & -12.5 & \textbf{1.06e-24} \\
March & 11.4 & 23.6 & 81.6 & 76.9 & -12.7 & \textbf{1.35e-30} \\
April & 14.6 & 21.4 & 93.3 & 87.17 & -15.22 & \textbf{1.24e-40} \\
May & 21 & 17.6 & 106.5 & 91.47 & -15.8 & \textbf{2.81e-42} \\
June & 24.5 & 25.6 & 92.2 & 78.8 & -16.66 & \textbf{6.36e-51} \\ 
Total, first six months & 19.3 & 23.37 & 91.36 & 81.67 & -35.58 & \textbf{5.64e-223} \\
    \end{tabular}
    \end{center}
\end{table}

\setlength{\tabcolsep}{0.9pt}
\small
\begin{table}
\begin{center}
 \caption{Time to acceptance for each two consecutive years in 2016-2020. Mean, std, \textit{T-Test  statistic} and p-value. Results in bold are statistically significant.}
    \label{tbl:acc_time_stats_history}
    \begin{tabular}{c|cc|cc|c|c}
    \toprule
\multirow{2}{*}{Years} &
    \multicolumn{2}{c}{First year} &
    \multicolumn{2}{c}{Second year} &
    \multirow{2}{*}{t-test statistic} &
    \multirow{2}{*}{p-value} \\
& Mean & std & Mean & std & & \\
   \midrule
2016-2017 & 77.27 & 77.11 & 77.32 & 82.2 & 0.016 & 0.49 \\
2017-2018 & 77.32 & 82.2 & 73.55 & 76.85 & -1.078 & 0.14 \\
2018-2019 & 73.55 & 76.85 & 75.79 & 77.47 & 0.72 & 0.23 \\
2019-2020 & 75.79 & 77.47 & 65.95 & 75.47 & -3.93 .8 & \textbf{4.21e-5} \\
2019-2020 non-COVID-19 & 75.79 & 77.47 & 91.35 & 81.67 & 5.56  & \textbf{1.44e-8} \\

    \end{tabular}
    \end{center}
\end{table}

\begin{figure}
	    \includegraphics[width=\textwidth]{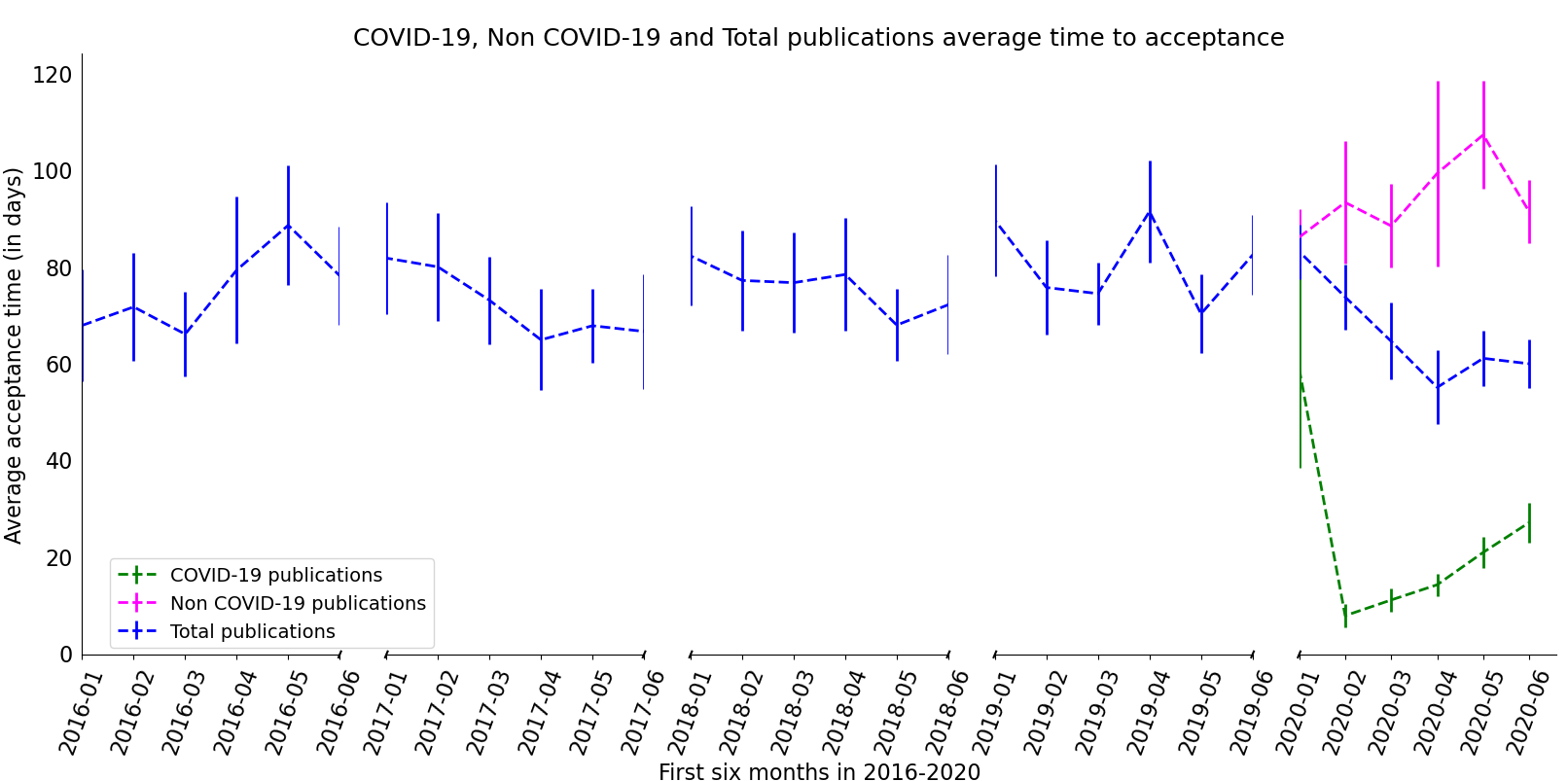}
        \caption{Mean time to acceptance in the examined Scopus journals for COVID-19 papers compared with non-COVID-19 and all published papers during the first six months of 2016-2020. The mean for each month is averaged over the number of journals (a ist of journals is shown in Table  \ref{tbl:cov_journals_growth}).} SEM is calculated and displayed separately for each month.
        \label{fig:cov-19-non_cov-total-acceptance_history}
\end{figure}

The results are also depicted in Figure \ref{fig:cov-19-non_cov-total-acceptance_history}, which displays the mean time to acceptance and the Standard Error of the Mean (SEM) averaged over all analysed journals from Table \ref{tbl:cov_journals_growth} in the first six months of each of the years 2016-2020. The figure also presents the mean and the SEM for COVID-19 and non-COVID-19 related publications in 2020. We observe an apparent trend for COVID-19 publications which first declines sharply from January to February, mainly attributed to the relatively low number of COVID-19 papers in January and thus the high SEM in January, and then inclines moderately from February onwards, mainly due to the increase of COVID-19 papers. However, despite the observed trend, COVID-19 publications seem to \say{enjoy} a shorter time to acceptance period compared both to non-COVID-19 publications in 2020 and to publications in 2016-2019. As we speculated before, our data and analyses show an association between non-COVID-19 and a longer time to acceptance. However, additional analysis is needed in order to further understand the impact of short time to acceptance for COVID-19 related papers both in relation to other, at the time of the pandemic, non-COVID-19 publications and to post-pandemic publications in general.


\subsection{Top Publishing Countries and International Collaboration}
In this section we focus on the source countries of the papers under analysis. We first analyse the countries with the highest number of publications and then proceed to international collaboration observing two facets as detailed in Section \ref{subsec:AnalysisApproach}. As before, we analyse these trends in the first six months of 2020 (for both COVID-19 and non-COVID-19 related papers) and the first six months of the previous four years.
\subsubsection{Top Publishing Countries}

\begin{figure}[ht]
	    \includegraphics[width=\textwidth]{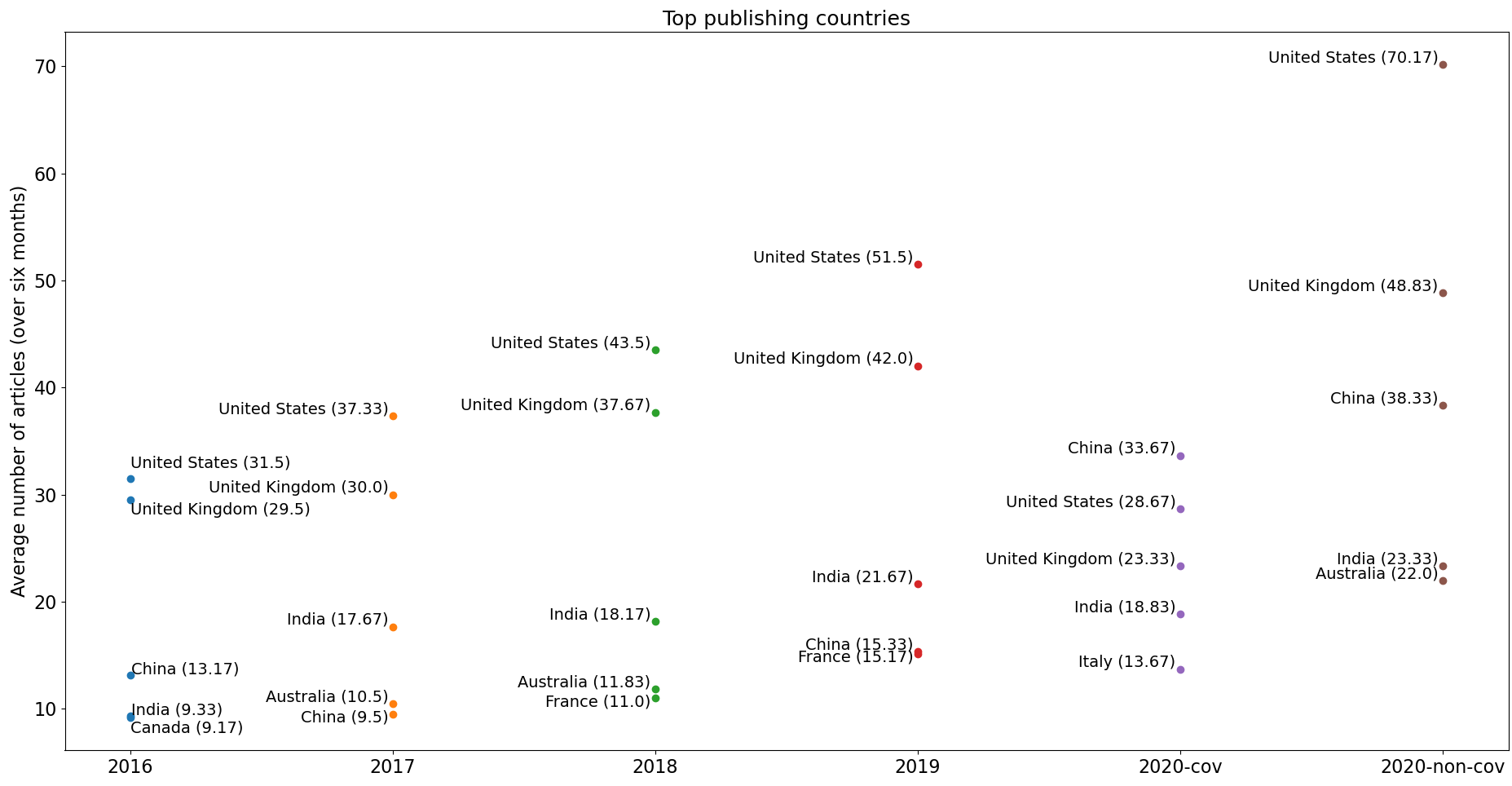}
        \caption{Top publishing countries for the Scopus selected journals in 2016-2020 and the number of publications by them. For each year the numbers are averaged over the first six months. 2020-cov and 2020-non-cov show the average number of COVID-19 and non-COVID-19 related publications respectively. (Numbers in parenthesis display the average number of papers).}
        \label{fig:top_pub_countries_history}
\end{figure}

We focus on the countries with the highest COVID-19 related publications and compare them to the top publishing countries of non-COVID-19 papers in the first six months of 2020, as well as to the historical data from the previous four years. In Table \ref{tbl:cov_countries} we report the comparison between COVID-19 and non-COVID-19 publications for the first six months of 2020. A longitudinal perspective over last five years is presented in Figure \ref{fig:top_pub_countries_history}. 
\\
\\
\\
\\
\\
\setlength{\tabcolsep}{0.3pt}
\newpage
\begin{center}
\small
\begin{longtable}[t]{cccc|cccc|cccc}
\caption[Covid-19 related top publishing countries]{Top COVID-19 and non-COVID-19 publishing countries and number of papers published by each country, in the first six months of 2020.}
\label{tbl:cov_countries} \\
\toprule
\multicolumn{4}{c}{January} &
    \multicolumn{4}{c}{February} &
    \multicolumn{4}{c}{March}\\
           Cov19 && Non-Cov19 &&  Cov19 && Non-Cov19 &&  Cov19 && Non-Cov19  \\
\midrule 
    Colombia &            2 &       United States &               64 &           China &           16 &      United Kingdom &               66 &           China &           33 &       United States &               60 \\
    Switzerland &            2 &      United Kingdom &               56 &  United Kingdom &            8 &       United States &               64 &   United States &           22 &      United Kingdom &               43 \\
         Canada &            2 &               China &               30 &       Hong Kong &            4 &               China &               32 &  United Kingdom &           13 &               China &               31 \\
      Hong Kong &            2 &           Australia &               22 &   United States &            4 &               India &               30 &          France &            6 &           Australia &               17 \\
          Spain &            2 &         Netherlands &               20 &        Colombia &            3 &           Australia &               25 &       Hong Kong &            6 &             Germany &               17 \\
        Ireland &            1 &              Canada &               18 &           Italy &            3 &               Italy &               24 &       Australia &            5 &              Brazil &               17 \\
          Nepal &            1 &         Switzerland &               17 &           Japan &            3 &             Germany &               23 &           Italy &            5 &         Switzerland &               16 \\
   Saudi Arabia &            1 &             Germany &               17 &           Nepal &            2 &              Canada &               20 &       Singapore &            5 &              Canada &               15 \\
        Nigeria &            1 &               Japan &               14 &       Argentina &            2 &         Netherlands &               19 &     Switzerland &            5 &               Italy &               14 \\
         Brazil &            1 &               India &               13 &           India &            2 &              France &               19 &          Norway &            3 &              France &               12 \\
\hline
\hline
    \multicolumn{4}{c}{April} &
    \multicolumn{4}{c}{May} &
    \multicolumn{4}{c}{June}\\
           Cov19 && Non-Cov19 &&  Cov19 && Non-Cov19 &&  Cov19 && Non-Cov19  \\
\midrule
                     China &           70  &              United States &               63  &              United States &           47  &       United States &               74  &   United States &           56  &       United States &               96  \\
             United States &           42  &             United Kingdom &               40  &             United Kingdom &           40  &               China &               36  &           China &           43  &               China &               64  \\
            United Kingdom &           36  &                      China &               37  &                      China &           39  &      United Kingdom &               30  &  United Kingdom &           42  &      United Kingdom &               58  \\
                     India &           34  &                      India &               25  &                      India &           33  &               India &               28  &           India &           41  &               India &               33  \\
                     Italy &           24  &                  Australia &               20  &                      Italy &           23  &               Italy &               19  &           Italy &           27  &           Australia &               32  \\
                    France &           13  &                     France &               18  &                     France &           16  &              France &               19  &          France &           15  &              France &               29  \\
 Iran &           11  &                      Spain &               18  &                    Germany &           13  &           Australia &               16  &         Germany &           12  &               Italy &               23  \\
                 Hong Kong &           11  &                Switzerland &               17  &  Iran &           13  &             Germany &               13  &           Japan &           11  &               Japan &               22  \\
                    Norway &           10  &                      Japan &               17  &                  Hong Kong &           13  &         Switzerland &               13  &          Brazil &           10  &               Spain &               22  \\
                    Canada &            9  &  Iran &               17  &                  Australia &           12  &              Sweden &               12  &          Canada &            8  &             Germany &               20  \\
\bottomrule
\end{longtable}
\end{center}


As can be seen from Figure \ref{fig:top_pub_countries_history}, while the top publishing countries have remained almost the same over the last five years, a consistent growth in the number of papers is evident. This is consistent with our previous findings, showing the total growth of publications over the years and most significantly in 2020.
It is interesting to note that despite the fact that the same countries are top publishing countries throughout the years regardless of the pandemic, Italy is the anomaly as it has never ranked in the top five before but during the pandemic has played a major part in COVID-19 research, as can be seen from its ranking as 5th in COVID-19 related publications. A similar pattern is displayed by Brazil and Hong Kong. Both are in the top 10 publishing countries for COVID-19 papers but with an average world ranking of 14 and 34 in the SJR country ranking, respectively. This can be explained by the large outbreak of the pandemic in these three countries during the examined months.
\\

\subsubsection{International Collaboration}
\paragraph{\textit{Diversity of international collaboration.}}
Recall from Section \ref{subsec:AnalysisApproach} that we measure diversity as the \textbf{number of countries} with which each country has collaborated over a given time period. We first analyse the impact of the COVID-19 pandemic on international collaborations at the time of the pandemic and then proceed to analyse how international collaboration has changed in diversity over the last five years.

\begin{figure}
	    \includegraphics[width=\textwidth]{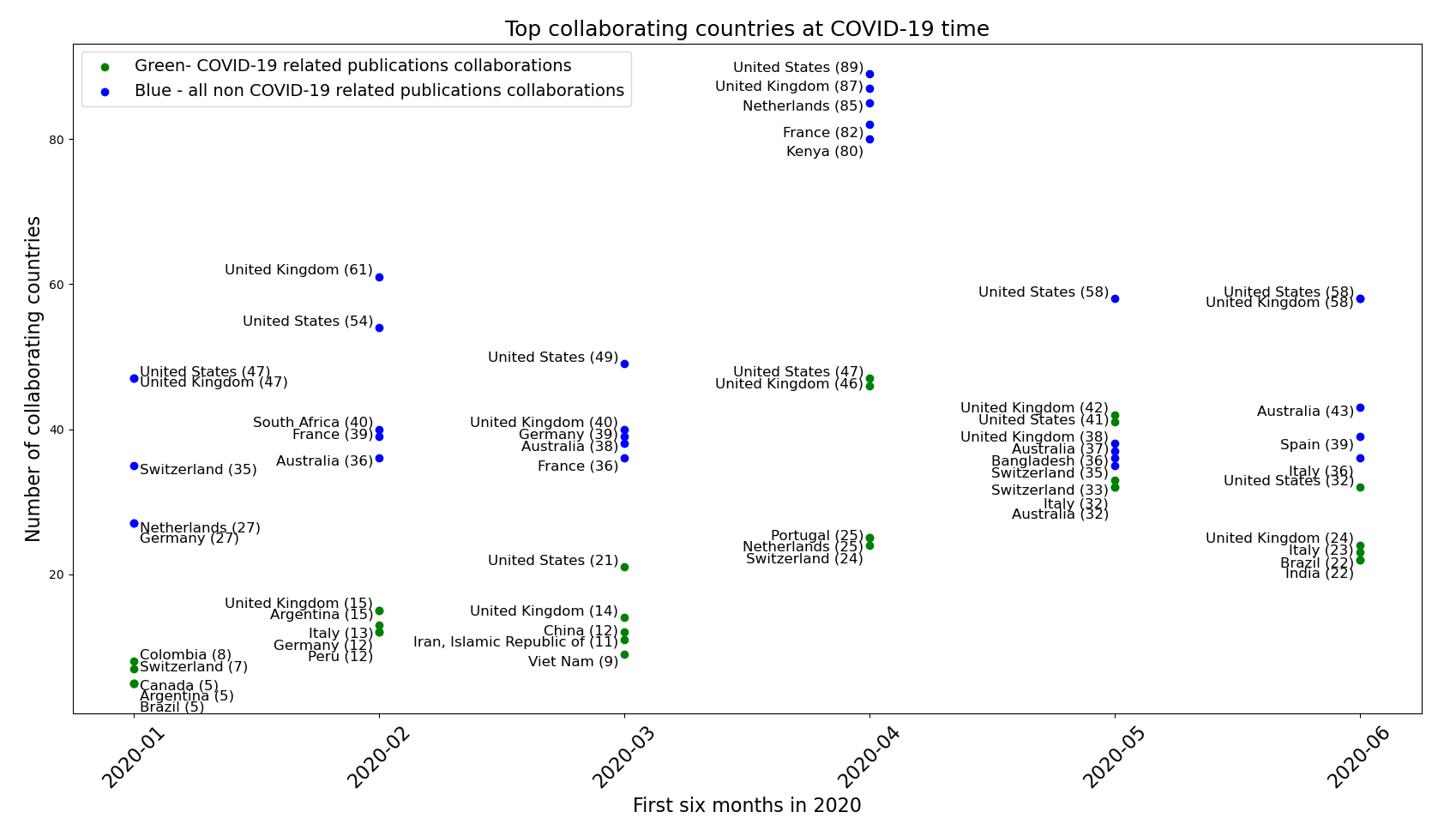}
        \caption{Top five diversely collaborative countries in the examined Scopus journals during the first six months of 2020 and the number of countries collaborating with them. Collaborating countries are shown for COVID-19 related publications (green) and non-COVID-19 related publications (blue). (Numbers in parentheses display the number of collaborating countries).}
        \label{fig:cov_19_top_collab_countries}
\end{figure}

Figure \ref{fig:cov_19_top_collab_countries} displays a comparison of COVID-19 and non-COVID-19 collaborations in 2020. For each country we observed the number of countries with whom it had collaborated for COVID-19 related papers and for non-COVID-19 related papers separately. This research was conducted separately for each of the months February-June of 2020\footnote{Data for January was insufficient to perform this test.} and also for the total first six months of 2020. The mean number of collaborating countries was calculated and a t-test was performed for each of these periods. The results are shown in Table \ref{tbl:collab_diversity_stats_cov_non_cov}. Our findings showed that for all periods tested, except for the month of May, the mean number of collaborating countries in non-COVID-19 papers was found to be statistically significantly greater than the mean number of collaborating countries for COVID-19 papers, with $p<0.05$.

\setlength{\tabcolsep}{0.9pt}
\small
\begin{table}
\begin{center}
 \caption{Number of collaborating countries for COVID-19 and non-COVID-19 papers in the first months of the pandemic. Mean, std, \textit{T-Test statistic} and p-value Results in bold are statistically significant.}
    \label{tbl:collab_diversity_stats_cov_non_cov}
    \begin{tabular}{c|cc|cc|c|c}
    \toprule
    \multirow{2}{*}{Months} &
    \multicolumn{2}{c}{COVID-19} &
    \multicolumn{2}{c}{non-COVID-19} &
    \multirow{2}{*}{t-test statistic} &
    \multirow{2}{*}{p-value} \\
& Mean & std & Mean & std & & \\
   \midrule
February & 6.32 & 3.59 & 13.65 & 12.88 & 4.72 & \textbf{3.8e-5} \\
March & 4.41 & 3.68 & 11.89 & 11.24 & 4.96 & \textbf{1.6e-6} \\
April & 10.45 & 9.31 & 48.5 & 28.6 & 12.43 & \textbf{2.46e-24} \\
May & 12.23 & 10.9 & 12.76 & 10.8 & 0.28 & 0.39 \\
June & 7.41 & 7.15 & 13.1 & 12.18 & 3.47 & \textbf{0.0004} \\ 
Total, first six months & 15.39 & 13.97 & 42.07 & 34.04 & -8.577 & \textbf{9.73e-16} \\
    \end{tabular}
    \end{center}
\end{table}

\setlength{\tabcolsep}{0.9pt}
\small
\begin{table}
\begin{center}
 \caption{Number of collaborating countries for each two consecutive years in 2016-2020. Mean, std, \textit{T-Test  statistic} and p-value. Results in bold are statistically significant.}
    \label{tbl:collab_diversity_stats_history}
    \begin{tabular}{c|cc|cc|c|c}
    \toprule
\multirow{2}{*}{Years} &
    \multicolumn{2}{c}{First year} &
    \multicolumn{2}{c}{Second year} &
    \multirow{2}{*}{t-test statistic} &
    \multirow{2}{*}{p-value} \\
& Mean & std & Mean & std & & \\
   \midrule
2016-2017 & 13.16 & 13.49 & 14.5 & 15.43 & 0.71 & 0.24 \\
2017-2018 & 14.5 & 15.43 & 22.32 & 18.54 & 3.63 & \textbf{0.0002} \\
2018-2019 & 22.32 & 18.54 & 30.81 & 24.24 & 3.2 & \textbf{0.0007} \\
2019-2020 & 30.81 & 24.24 & 42.94 & 34.62 & 3.49  & \textbf{0.0002} \\
2019-2020 COVID-19 & 30.81 & 24.24 & 15.39 & 13.97 & -6.13  & \textbf{-2e-9} \\
2019-2020 non-COVID-19& 30.81 & 24.24 & 42.07 & 34.04 & 3.233  & \textbf{0.00069} \\

    \end{tabular}
    \end{center}
\end{table}

\begin{figure}
	    \includegraphics[width=\textwidth]{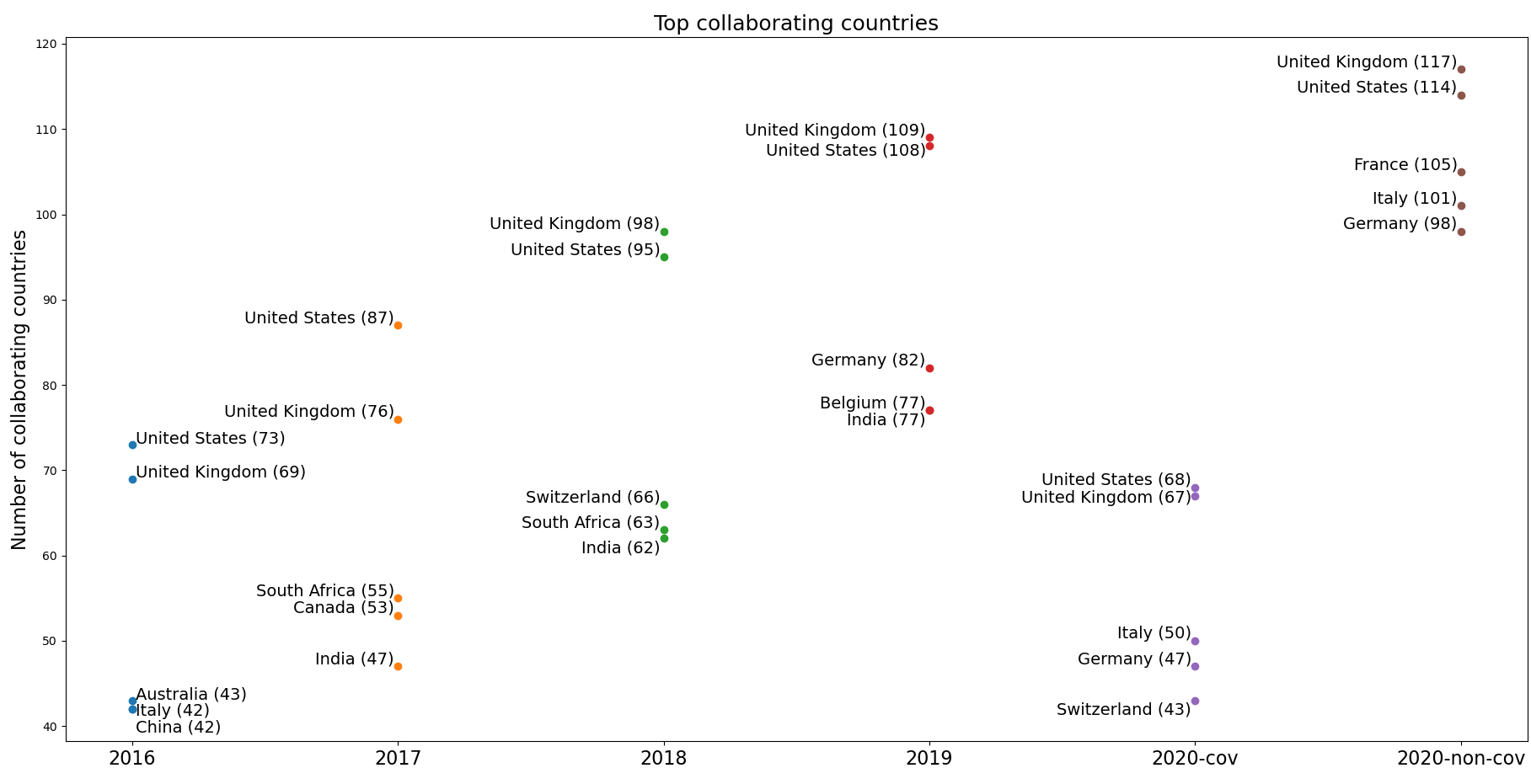}
        \caption{Top five diversely collaborative countries in the examined Scopus journals in 2016-2020 and the number of countries collaborating with them. For each year the numbers are aggregated over the first six months (repeating countries were removed). 2020-cov and 2020-non-cov show the top five collaborative countries for COVID-19 and non-COVID-19 related publications, respectively. (Numbers in parentheses display the number of collaborating countries).}
        \label{fig:top_collab_countries_history}
\end{figure}

Following the short term analysis, we conducted a longitudinal one observing the years 2016-2020. Figure \ref{fig:top_collab_countries_history} presents the international collaboration diversity in this perspective. For each country we observed the aggregated number of countries with which it had collaborated in the first six months of each year (repeated countries were removed). Similar to the trend we saw for top publishing countries, the top collaborating countries have remained almost the same over the last five years and a consistent growth in the number of countries with which each of these top countries collaborated is evident. 
We performed pairwise t-tests for each two consecutive years in our study. The test included all countries which collaborated with at least one other country in the first six months of the years 2016-2020. Results are shown in Table \ref{tbl:collab_diversity_stats_history}.
The results show that for all examined periods, the number of collaborating countries is steadily increasing. The differences are statistically significant from the 2017-2018 period onward, $p<0.05$. In addition, we performed t-tests comparing international collaboration on COVID-19 papers in 2019 with those in 2020 as well as non-COVID-19 papers in 2019 with those in 2020. While both showed a statistically significant difference, the direction is reversed as the mean number of collaborating countries for 2020 COVID-19 papers is statistically significantly \textbf{smaller} than that in 2019, but the mean number of collaborating countries for 2020 non-COVID-19 papers is statistically significantly \textbf{larger} than that in 2019. $p<0.05$ for all accounts.

An increase in collaboration diversity over the years has been demonstrated in previous works as well \citep{leydesdorff2008international, wagner2005network}. The results presented here complement these findings by showing that the year 2020 has experienced the largest increase in the number of collaborating counties.
However, the findings from the comparison of 2019 to the 2020 COVID-19 papers along with the findings when comparing COVID-19 and non-COVID-19 collaborations (as shown in Table \ref{tbl:collab_diversity_stats_cov_non_cov}) show that collaboration in COVID-19 papers is, surprisingly, low. Specifically, it is lower when compared to non-COVID-19 papers, lower when compared to collaboration in the past and lower than what we would expect during a pandemic, where collaboration is of increased significance.


\paragraph{\textit{Volume of international collaboration}}
Recall from Section \ref{subsec:AnalysisApproach} that we measure \say{volume} as the \textbf{number of publications} which each country has published in collaboration with other countries over a given time period. For this analysis we performed two statistical tests for both the short-term COVID-19 pandemic months and for the long-term 2016-2020 period: $\chi^2$ test \citep{pearson1900x} and \textit{t-test}.

\begin{figure}
	    \includegraphics[width=\textwidth]{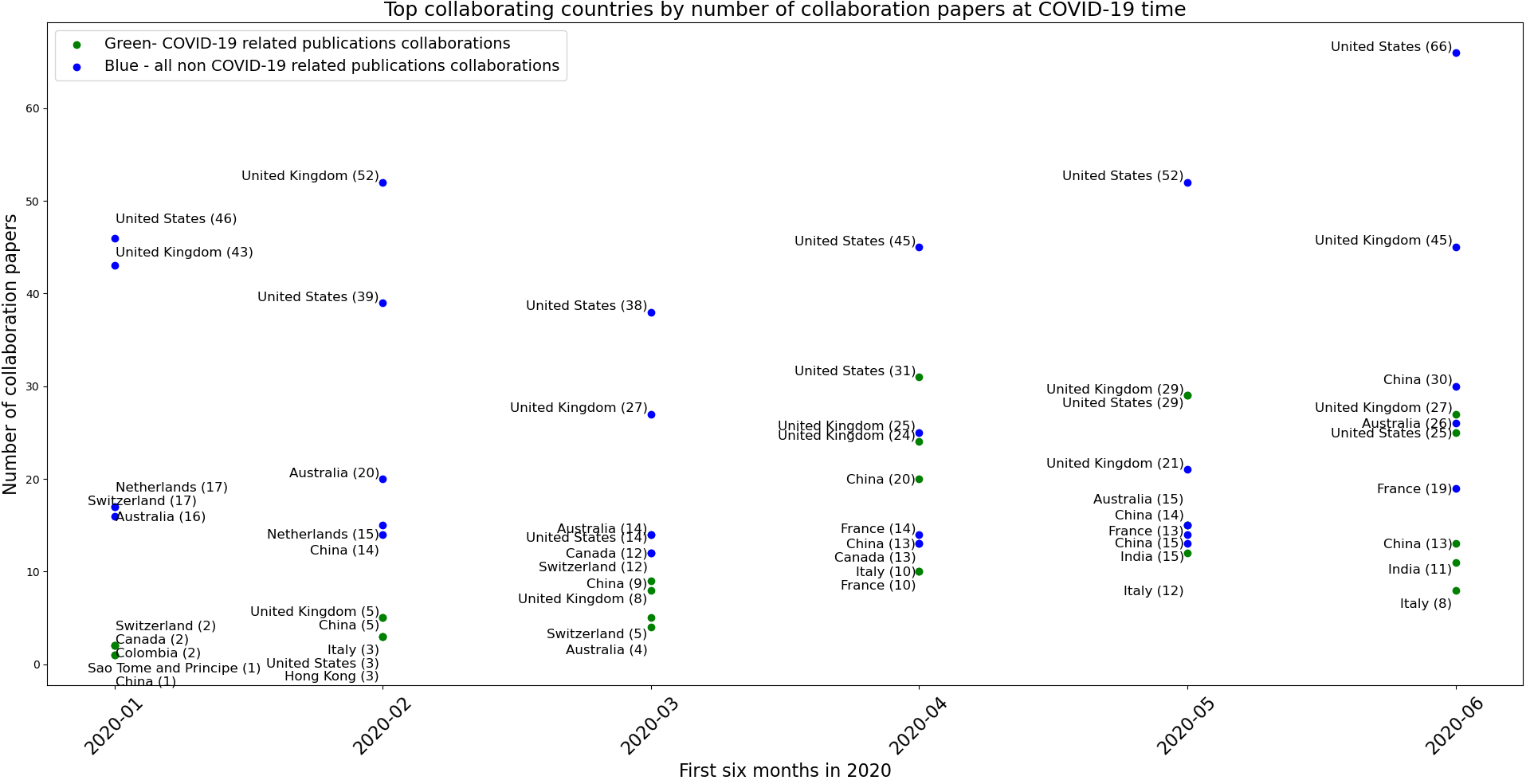}
        \caption{Top five collaborative countries by volume in the examined Scopus journals during the first six months of 2020 and the number of collaborative publications. The number of collaborative publications is shown for COVID-19 related publications (green) and non-COVID-19 related publications (blue), (Numbers in parentheses display the number of collaborative papers).}
        \label{fig:cov_19_top_collab_countries_num_papers}
\end{figure}

\begin{figure}
	    \includegraphics[width=\textwidth]{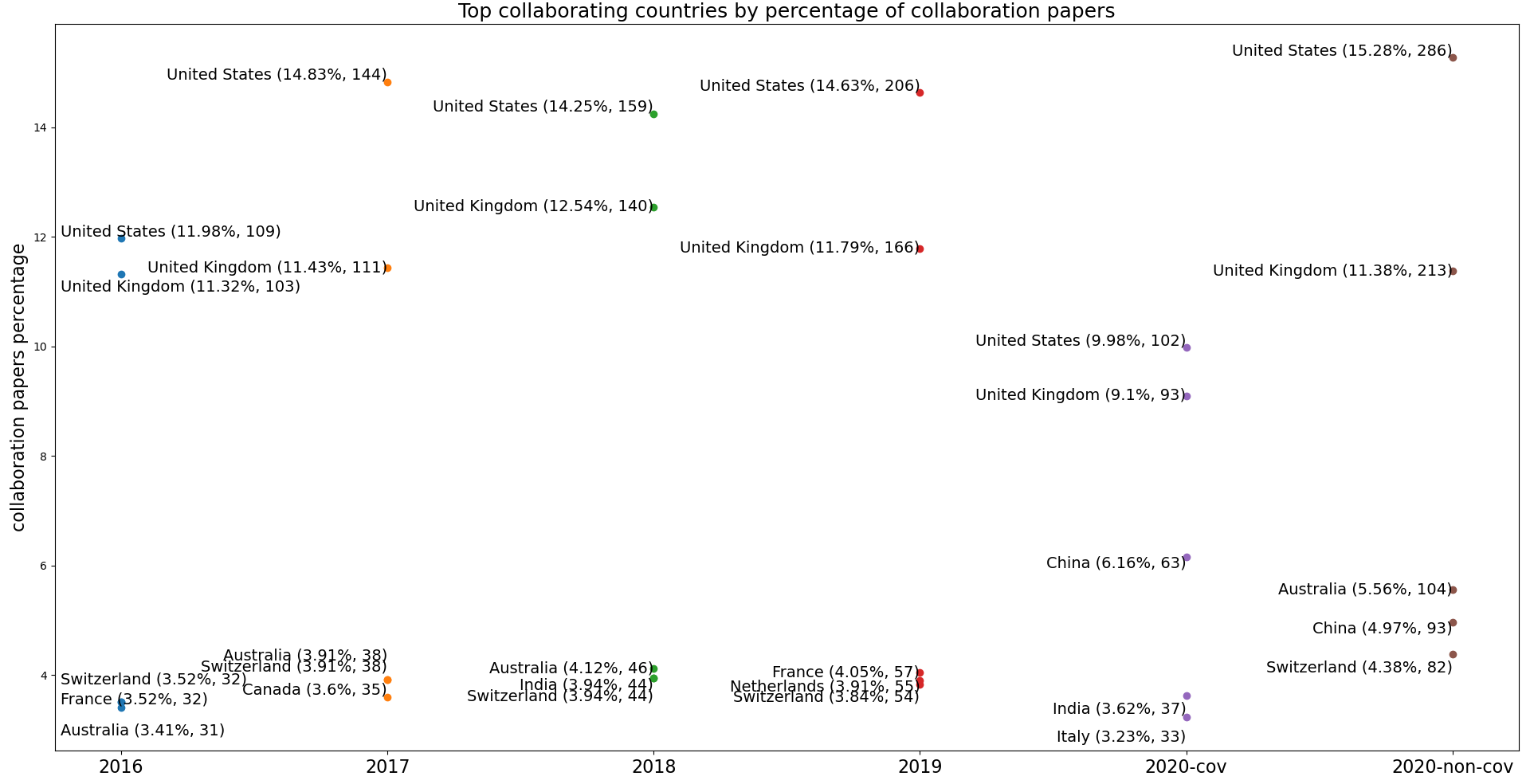}
        \caption{Top five collaborative countries by volume in the examined Scopus journals in 2016-2020 and the percentage of collaborative publications. 2020-cov and 2020-non-cov show the top five collaborative countries for COVID-19 and non-COVID-19 related publications, respectively. Percentage calculated is the number of collaborative papers relative to all of the papers published in the same time period by the same country. (Numbers in parentheses display the percentage and the number of collaborative papers).}
        \label{fig:top_collab_countries_num_papers_history}
\end{figure}

Figure \ref{fig:cov_19_top_collab_countries_num_papers} displays the countries with the highest number of papers written in international collaboration for COVID-19 papers compared with non-COVID-19 papers in the first six months of 2020. Figure \ref{fig:top_collab_countries_num_papers_history} displays the countries with the highest number of papers written in international collaboration over the last five years. We can observe two interesting collaboration patterns from these figures. The first is that the number of papers written in collaboration has continually increased over the last five years, for the top collaborating countries. This is consistent with our findings of growth in total publications over the last five years, as can be seen in Figure \ref{fig:Publication_growth_cov_totals}. The second finding that can be observed is that although the US and the UK have remained the top two collaborating countries, when observing collaboration for COVID-19 publications, China is extremely collaborative and Italy is in the top 5 collaborating countries. This can be explained by the pandemic originating in China and its wide spread in Italy.

For the $\chi^2$ test we define two extreme cases, papers written with no collaboration at all, and papers written in any form of collaboration, meaning at least two countries collaborated in authorship of that paper. Thus we compare single country authored papers to multi-country authored papers.
The test was performed for each of the months February-June of 2020\footnote{As before, data for January was insufficient to perform this test.} separately and as a sum over the first six months of 2020. In each time period we examined how many papers were authored by a single country (all authors from the same country) and how many papers were authored in collaboration with other countries. This was done for COVID-19 papers and for non-COVID-19 papers.
The results are shown in Table \ref{tbl:collab_num_papers_stats_chi_cov_non_cov}. 
Based on the $\chi^2$ statistic and $p$ values we can conclude that for the months of April, May and June 2020 and for the total over the six months, the type of paper (i.e., COVID-19 or non-COVID-19 related) is significantly associated with the authorship by a single country or multiple countries. Specifically, authorship by a single country is indicative of COVID-19 related papers while co-authorship by multiple countries is indicative of non-COVID-19 related papers.

\setlength{\tabcolsep}{0.9pt}
\small
\begin{table}
\begin{center}
 \caption{Number of papers published by a single country and by multiple countries for COVID-19 and non-COVID-19 papers in the first months of the pandemic,  \textit{$\chi^2$ statistic} and p-value. Results in bold are statistically significant.}
    \label{tbl:collab_num_papers_stats_chi_cov_non_cov}
    \begin{tabular}{c|cc|cc|c|c}
    \toprule
    \multirow{2}{*}{Months} &
    \multicolumn{2}{c}{COVID-19} &
    \multicolumn{2}{c}{non-COVID-19} &
    \multirow{2}{*}{$\chi^2$ statistic} &
    \multirow{2}{*}{p-value} \\
& Single Country & Multi Country & Single Country & Multi Country & & \\
   \midrule
February & 18 & 67 & 187 & 418 & 2.93 & 0.08 \\
March & 61 & 85 & 151 & 281 & 1.9 & 0.16 \\
April & 188 & 248 & 185 & 457 & 22.8 & \textbf{1.75e-6} \\
May & 181 & 251 & 173 & 352 & 7.76 & \textbf{0.005} \\
June & 226 & 204 & 235 & 510 & 49.6 & \textbf{1.86e-12} \\ 
Total, first six months & 681 & 870 & 1077 & 2394 & 77.58 & \textbf{1.27e-18} \\
    \end{tabular}
    \end{center}
\end{table}

\setlength{\tabcolsep}{0.9pt}
\small
\begin{table}
\begin{center}
 \caption{Number of papers published by a single country and multiple countries for each two consecutive years in 2016-2020, \textit{$\chi^2$ statistic} and p-value. Results in bold are statistically significant.}
    \label{tbl:collab_num_papers_stats_chi_history}
    \begin{tabular}{c|cc|cc|c|c}
    \toprule
\multirow{2}{*}{Years} &
    \multicolumn{2}{c}{First year} &
    \multicolumn{2}{c}{Second year} &
    \multirow{2}{*}{$\chi^2$ statistic} &
    \multirow{2}{*}{p-value} \\
& Single Country & Multi Country & Single Country & Multi Country & & \\
   \midrule
2016-2017 & 582 & 909 & 616 & 954 & 0.006 & 0.94 \\
2017-2018 & 616 & 954 & 708 & 1232 & 2.66 & 0.1 \\
2018-2019 & 708 & 1232 & 854 & 1602 & 1.33 & 0.25 \\
2019-2020 & 854 & 1602 & 1756 & 3260 & 0.03  & 0.96 \\
2019-2020 COVID-19 & 854 & 1602 & 681 & 870 & 33.18  & \textbf{8.38e-9} \\
2019-2020 non-COVID-19 & 854 & 1602 & 1077 & 2394 & 9  & \textbf{0.003} \\
    \end{tabular}
    \end{center}
\end{table}
Additional $\chi^2$ tests were performed to analyse the longitudinal aspect of our study. Specifically, has the pandemic affected the number of papers internationally co-authored compared to previous years? In this test we measured the mean number of papers authored by single countries and the mean number of papers authored by multiple countries in the first six months of each two consecutive years in our study, comparing 2016-2017, 2017-2018, 2018-2019, 2019-2020. This can be seen in Table \ref{tbl:collab_num_papers_stats_chi_history}.
Our results for the longitudinal $\chi^2$ test are statistically significant at $p<0.05$ in two specific cases: when comparing international co-authorship for 2019 papers to 2020 COVID-19 papers and when comparing authorship for 2019 papers to 2020 non-COVID-19 papers.
Taken jointly, the results from the short-term and the longitudinal $\chi^2$ tests indicate that for COVID-19 related papers, the volume of papers co-authored by multiple countries is \textbf{low}, both in comparison to non-COVID-19 related papers and to previous years' international collaboration behaviour.
In order to better understand the findings from our $\chi^2$ tests, we performed a series of short-term analysis t-tests and a similar series of long-term analysis t-tests.
In the short-term analysis we measured, for every country, the number of papers written in international collaboration. This was performed in each of the months February-June of 2020\footnote{Data for January was insufficient to perform this test.} and for the total first six months of 2020 for COVID-19 related papers and for non-COVID-19 related ones. The mean number of collaborative papers was calculated and a t-test was performed for each of these periods. 
The results are displayed in Table \ref{tbl:collab_num_papers_stats_cov_non_cov} and show that for the examined periods of February, March and June and the total first six months of 2020, the mean number of COVID-19 collaborative papers is statistically significantly smaller than for non-COVID-19, with $p<0.05$.

\setlength{\tabcolsep}{0.9pt}
\small
\begin{table}
\begin{center}
 \caption{Number of internationally collaborated papers for COVID-19 and non-COVID-19 papers in the first months of the pandemic. Mean, std, \textit{T-Test statistic} and p-value. Results in bold are statistically significant.}
    \label{tbl:collab_num_papers_stats_cov_non_cov}
    \begin{tabular}{c|cc|cc|c|c}
    \toprule
    \multirow{2}{*}{Months} &
    \multicolumn{2}{c}{COVID-19} &
    \multicolumn{2}{c}{non-COVID-19} &
    \multirow{2}{*}{t-test statistic} &
    \multirow{2}{*}{p-value} \\
& Mean & std & Mean & std & & \\
   \midrule
February & 1.55 & 1.02 & 5.16 & 7.76 & 4.08 & \textbf{5e-5} \\
March & 2.5 & 2.76 & 4 & 5.9 & 1.76 & \textbf{0.04} \\
April & 3.75 & 5.34 & 4.35 & 5.36 & 0.7 & 0.24 \\
May & 4.18 & 5.67 & 3.96 & 6.3 & 0.22 & 0.4 \\
June & 3.46 & 4.91 & 5.93 & 9.53 & 2.03 & \textbf{0.022} \\ 
Total, first six months & 8.45 & 15.64 & 15.55 & 32.9 & 2.3 & \textbf{0.011} \\
    \end{tabular}
    \end{center}
\end{table}

\setlength{\tabcolsep}{0.9pt}
\small
\begin{table}
\begin{center}
 \caption{Number of internationally collaborated papers for each two consecutive years in 2016-2020. Mean, std, \textit{T-Test  statistic} and p-value. Results in bold are statistically significant.}
    \label{tbl:collab_num_papers_stats_history}
    \begin{tabular}{c|cc|cc|c|c}
    \toprule
\multirow{2}{*}{Years} &
    \multicolumn{2}{c}{First year} &
    \multicolumn{2}{c}{Second year} &
    \multirow{2}{*}{t-test statistic} &
    \multirow{2}{*}{p-value} \\
& Mean & std & Mean & std & & \\
   \midrule
2016-2017 & 7.77 & 15.06 & 7.88 & 17.55 & 0.05 & 0.48 \\
2017-2018 & 7.88 & 17.55 & 9.41 & 20.16 & 0.63 & 0.26 \\
2018-2019 & 9.41 & 20.16 & 11.87 & 24.57 & 0.89 & 0.19 \\
2019-2020 & 11.87 & 24.57 & 20.5 & 45.13 & 2.07  & \textbf{0.019} \\
2019-2020 COVID-19 & 11.87 & 24.57 & 8.45 & 15.64 & -1.3  & 0.09 \\
2019-2020 non-COVID-19 & 11.87 & 24.57 & 15.55 & 32.93 & 1.08  & 0.14 \\
    \end{tabular}
    \end{center}
\end{table}

Following the short-term analysis we conducted a longitudinal one, observing the years 2016-2020. For every country, we observed the total number of papers written in international collaboration in the first six months of each year.
We performed pairwise t-tests for each two consecutive years in our study. 
From Table \ref{tbl:collab_num_papers_stats_history} we can observe that a statistically significant difference exists only for the period 2019-2020. The mean number of internationally collaborated papers in 2019 is statistically significantly smaller than the mean number of internationally collaborated papers in 2020,  with $p<0.05$. However, this difference is not statistically significant when comparing 2019 collaborated papers separately to 2020 COVID-19 papers and to 2020 non-COVID-19 papers.
These findings indicate that when international collaboration is measured by the volume of collaborated papers, the COVID-19 pandemic does not seem to affect the increase in international collaboration.

\section{Conclusion and Discussion}
\label{sec:Conclusion}

In this study we have analysed how the COVID-19 pandemic effected the publication patterns in biomedical literature.  We employed two types of analyses to address each of our research questions - short-term analysis and a longitudinal analysis of preprint servers and peer reviewed journals. 

Our analysis showed a significant increase in published papers both in peer reviewed journals and in preprint servers compared to previous years. The new MedRxiv preprint server especially stands out with an exceptionally large increase in publications and we expect this preprint server to continue this pattern post-pandemic.
Notably, while the increase in publication in preprint servers has occurred for both COVID-19 and non-COVID-19 related papers, this is not the case for the journals we have analysed. In these journals virtually the entire growth was due to COVID-19 papers while, on average, the volume of non-COVID-19 papers has remained similar to previous years.
Our results also showed that high ranked journals publish more COVID-19 papers than low ranked journals.
It is further apparent that journals had responded quickly to the pandemic by lowering the time to acceptance for COVID-19 papers. Unfortunately, this seems to have come at a cost for non-COVID-19 papers, whose time to acceptance was longer than that which was observed in previous years (and obviously longer than that of COVID-19 papers). While our analysis suggests strong supportive evidence to that conclusion, one cannot definitively rule out other \say{hidden} contributors which are outside the scope of this work.

Taken jointly, the non-increasing volume and longer time to acceptance of non-COVID-19 papers may lead to a slow down in non-COVID-19 related research and publication, at the very least in journal publications. In future work, we intend to extend our analysis  into these findings in order to further understand if a \say{slow down} can be observed.
While such research may not be \say{urgent}, it is, presumably, of no less importance than the current crisis. 
On the other hand, we do observe an increase in the use of preprint servers \textit{irrespective of COVID-19 publications}. This may indicate that the community is recognizing the above phenomena and adjusting their publication behaviour accordingly. 

Turning our focus to the countries authoring these papers, we observed the top publishing countries and international collaboration patterns. We observed that while the US and the UK have remained the top publishing and collaborating countries, Italy, Brazil and Hong Kong have produced a significant amount of COVID-19 related papers as well, disproportional to their lower ranking in respect to the number of non-COVID-19 papers authored in these countries over the previous four years. This observation could be explained by the major impact this pandemic has had on these countries.
Our results further showed consistent growth over the last five years in international collaboration when examining both collaboration diversity and volume. 
During the pandemic, we observed that the volume of COVID-19 related papers written in collaboration had increased compared to non-COVID-19 papers.
However, contrary to our original hypothesis, international collaboration diversity in COVID-19 papers was lower than in non-COVID-19 papers and lower than previous years. We observed that most COVID-19 papers were authored by a single country or only very few countries. This can be explained by the complexity of conducting international studies during the pandemic. However, this may also suggest that countries are doing a significant amount of COVID-19 research nationally and knowledge is officially shared only after the research has been published. 
This phenomenon is obviously undesired, especially at a time of international crisis.

We recognize that the current study is limited by the amount, quality and diversity of the data used. In the context of this work, the number of COVID-19 related papers was relatively low especially in the first 3 months of 2020. This could skew our findings in respect to both time to acceptance and collaboration analysis. In addition, we chose to focus on biomedical publications alone. Observing additional fields could lead to a broader trend which may not necessarily align with our results. Additional potential limitations to our study relate to our selection of data sources. We have selected only three public repositories to focus on due to the large increase of papers in these archives as well as their relevance to the selected fields. Similarly, we focused only on journals indexed by Scopus. A wider perspective on the matter could be obtained by including additional repositories as well as data from other indexing sources such as Web Of Science and Microsoft Academic. While this work is, to the best of our knowledge, the most extensive one on all four accounts when discussing scholarly publications during the COVID-19 outbreak, a larger analysis may reveal additional or other trends which were not captured here. 
We plan to extend this work further in several directions: First, we plan to apply more advanced analysis techniques and statistical methods such as time series and unsupervised learning methods \citep{madsen2007time, han2011data} and mixed effects modeling \citep{laird1982random} to our data. This could assist in identifying additional publication trends which were not revealed in the current study. Second, we wish to investigate the  long-term effects of this pandemic on scholarly research, both in the biomedical literature as well as in other fields. Such research would complement this work by analysing (hopefully) post-pandemic changes in citation, collaboration, time to acceptance patterns and additional scholarly publication trends.

\newpage

\section{References}
\printbibliography[heading=none]





\end{document}